\documentclass{l4dc2025}

    \usepackage{PaperSty}
    \usepackage{mathtools}
    \usepackage{amsfonts}
    \usepackage{amssymb,latexsym}
    \usepackage[psamsfonts]{eucal}
    \usepackage{indentfirst}
    \usepackage{setspace}
    \usepackage{threeparttable}
    \usepackage{float}
    \usepackage{afterpage}
    \usepackage{placeins}
    \usepackage{cancel}
    \usepackage{leftidx}
    \usepackage{breqn}
    \usepackage[export]{adjustbox}
    \usepackage{comment}
    \usepackage{soul}
    \usepackage{siunitx}
    \usepackage{subcaption}
    \usepackage{mwe}
    \usepackage{changepage}
    \usepackage{enumitem}
    \usepackage{multicol}
    \PassOptionsToPackage{ruled, linenumbered, lined}{algorithm2e}
    \usepackage{subcaption}
    \usepackage{wrapfig}
    \usepackage{sidecap}
    \usepackage{hyperref}
    \usepackage{cleveref}
    \usepackage[cal=cm]{mathalpha}
    \usepackage{aliascnt}
    \usepackage{paracol}

\hypersetup{
    colorlinks=true,
    urlcolor=blue
}

    \let\originalleft\left
    \let\originalright\right
    \renewcommand{\left}{\mathopen{}\mathclose\bgroup\originalleft}
    \renewcommand{\right}{\aftergroup\egroup\originalright}

    \graphicspath{ {Figures/} }
    \DeclareGraphicsExtensions{.png}

\ifjmlrcleveref
    \newaliascnt{atheorem}{theorem}
    \newtheorem{atheorem}[atheorem]{Theorem}
    \aliascntresetthe{atheorem}
    \crefname{atheorem}{theorem}{theorems}

    \newaliascnt{aexample}{theorem}
    \newtheorem{aexample}[aexample]{Example}
    \aliascntresetthe{aexample}
    \crefname{aexample}{example}{examples}

    \newaliascnt{alemma}{theorem}
    \newtheorem{alemma}[alemma]{Lemma}
    \aliascntresetthe{alemma}
    \crefname{alemma}{lemma}{lemmas}

    \newaliascnt{aproposition}{theorem}
    \newtheorem{aproposition}[aproposition]{Proposition}
    \aliascntresetthe{aproposition}
    \crefname{aproposition}{proposition}{propositions}

    \newaliascnt{aremark}{theorem}
    \newtheorem{aremark}[aremark]{Remark}
    \aliascntresetthe{aremark}
    \crefname{aremark}{remark}{remarks}

    \newaliascnt{acorollary}{theorem}
    \newtheorem{acorollary}[acorollary]{Corollary}
    \aliascntresetthe{acorollary}
    \crefname{acorollary}{corollary}{corollaries}

    \newaliascnt{adefinition}{theorem}
    \newtheorem{adefinition}[definition]{Definition}
    \aliascntresetthe{adefinition}
    \crefname{adefinition}{definition}{definitions}

    \newaliascnt{aconjecture}{theorem}
    \newtheorem{aconjecture}[aconjecture]{Conjecture}
    \aliascntresetthe{aconjecture}
    \crefname{aconjecture}{conjecture}{conjectures}

    \newaliascnt{afact}{theorem}
    \newtheorem{afact}[afact]{Fact}
    \aliascntresetthe{afact}
    \crefname{afact}{fact}{facts}

    \newaliascnt{aassumption}{theorem}
    \newtheorem{aassumption}[aassumption]{Assumption}
    \aliascntresetthe{aassumption}
    \crefname{aassumption}{assumption}{assumptions}

    \newaliascnt{aaxiom}{theorem}
    \newtheorem{aaxiom}[aaxiom]{Axiom}
    \aliascntresetthe{aaxiom}
    \crefname{aaxiom}{axiom}{axioms}
\else
    \newtheorem{atheorem}{Theorem}

    \newtheorem{aproposition}{Proposition}
    \newtheorem{aremark}{Remark}
    
    \newtheorem{adefinition}{Definition}

    \newtheorem{aassumption}{Assumption}
    
\fi

\newenvironment{aproof}[1][Proof]{%
  \par\noindent\textit{#1.} \rmfamily
}{%
  \hfill\ensuremath{\blacksquare}\par
}
\crefname{algorithm}{Algorithm}{Algorithms}

\usepackage{xcolor}
\usepackage{enumitem}

\title[RLBUS: RL Backup Shield]{Safe Exploration in Reinforcement Learning: Training Backup Control Barrier Functions with Zero Training-Time Safety Violations}

\usepackage{times}

\coltauthor{\Name{Pedram Rabiee} \Email{pedram.rabiee@uky.edu}\\
 \Name{Amirsaeid Safari} \Email{amirsaeid.safari@uky.edu}\\
 \addr Department of Mechanical and Aerospace Engineering, University of Kentucky}

\begin{document}

\maketitle

\begin{abstract}
This paper introduces the reinforcement learning backup shield (RLBUS), an algorithm that guarantees safe exploration in reinforcement learning (RL) by incorporating backup control barrier functions (BCBFs). RLBUS constructs an implicit control forward invariant subset of the safe set using multiple backup policies, ensuring safety in the presence of input constraints. 
While traditional BCBFs often result in conservative control forward-invariant sets due to the design of backup controllers, RLBUS addresses this limitation by leveraging model-free RL to train an additional backup policy, which enlarges the identified control forward invariant subset of the safe set. This approach enables the exploration of larger regions in the state space with zero safety violations during training. The effectiveness of RLBUS is demonstrated on an inverted pendulum example, where the expanded invariant set allows for safe exploration over a broader state space, enhancing performance without compromising safety.\footnote{\color{blue}\href{https://github.com/pedramrabiee/safe_rl}{\textcolor{blue}{The source code of this work is available at: https://github.com/pedramrabiee/safe\_rl}}}

\end{abstract}


\section{Introduction}

Safe reinforcement learning (RL) has emerged as a key area of research for deploying RL to real-world applications while ensuring safe interactions with the environment. Many existing approaches in safe RL aim to develop a safe policy by the end of training \citep{haarnoja2018soft, chow2018risk}; however, these methods often fall short in ensuring safety throughout the training process. More recent methods have focused on reducing unsafe interactions during learning \citep{schreiter2015safe, wachi2018safe, achiam2017constrained}, yet achieving full safety guarantees over the entire training horizon remains challenging.

Model-based approaches have integrated Lyapunov stability analysis to enhance safety during online exploration \citep{koller2018learning, berkenkamp2017safe, wang2018safe}. Despite effectively constraining system behavior, these methods lack a unified framework to simultaneously ensure safety and optimize performance. While some researchers have proposed switching between predefined backup controllers to ensure safety \citep{alshiekh2018safe, wabersich2018linear, bastani2021safe}, these approaches rely on statistical guarantees that accumulate errors over time, limiting their effectiveness for systems requiring long-term or infinite-horizon safety assurances.

A common approach to ensuring safety over an infinite horizon involves determining a policy that renders a specified safe set forward invariant \citep{blanchini1999set}. However, achieving controlled invariance of the safe set is challenging, particularly under input constraints, which often prevent finding control inputs that preserve the forward invariance of the entire set. Methods like Hamilton-Jacobi reachability analysis \citep{mitchell2005time} and sum-of-squares programming \citep{korda2014convex} provide computational approaches for determining a control forward-invariant subset of the safe set, but they can be computationally intensive for high-dimensional systems. Control barrier functions (CBFs) offer an alternative to traditional offline verification methods by constructing control-invariant sets that ensure online safety \citep{wieland2007constructive, ames2016control, rabiee2023composition, safari2023time}. However, designing CBFs that account for input constraints remains a challenging task.

% \begin{wrapfigure}{l}{0.5\textwidth}
\begin{figure}
    \centering
    \includegraphics[width=0.5\linewidth]{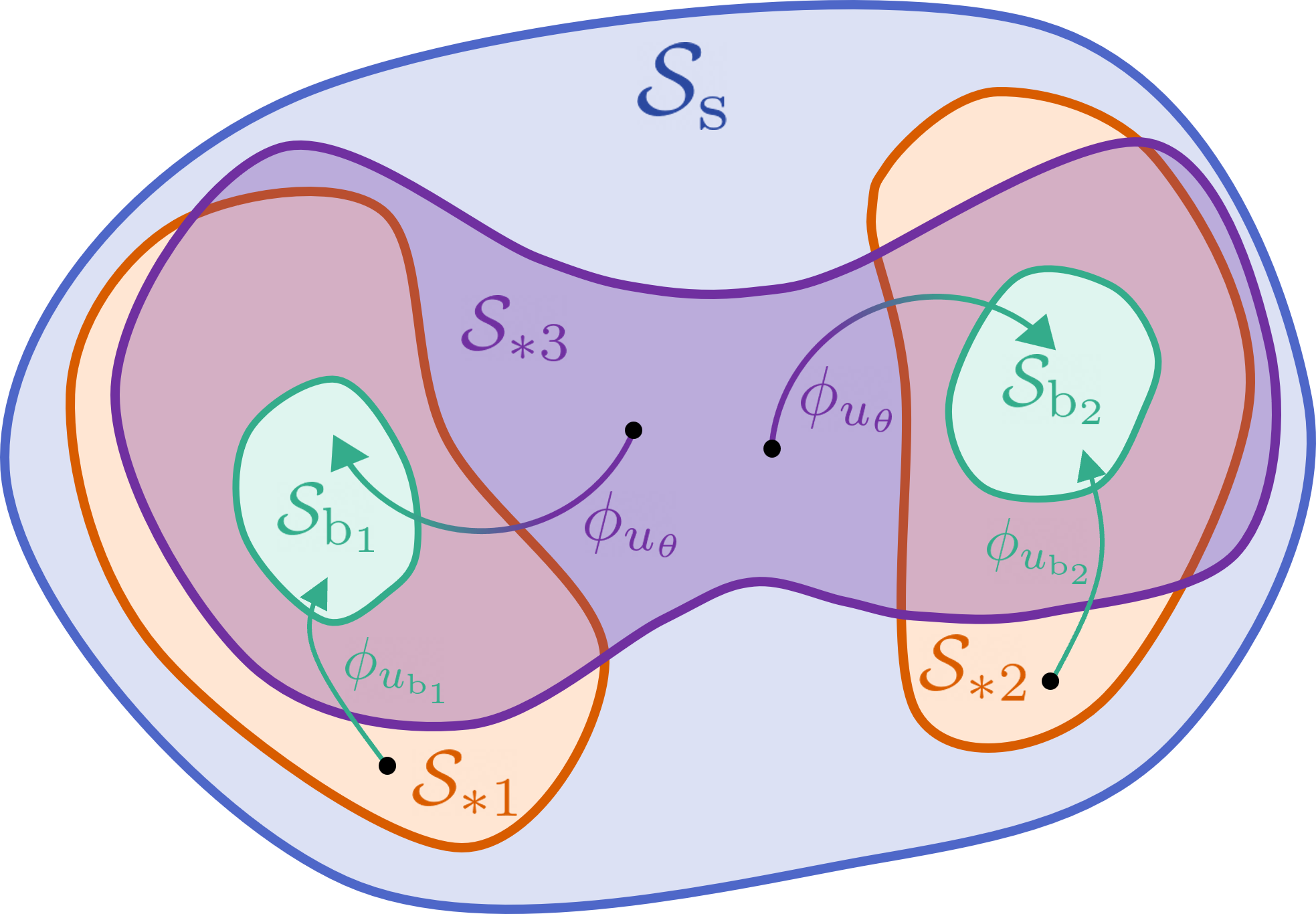}
\caption{The schematic illustrates the safe set $\SSS_\rms$, backup sets $\SSS_{\rmb_1}$, $\SSS_{\rmb_2}$, and forward invariant sets $\SSS_{*1}$, $\SSS_{*2}$ which are the finite-time safe backward images of $\SSS_{\rmb_1}$, $\SSS_{\rmb_2}$ under the user-designed backup policies $u_{\rmb_1}$ and $u_{\rmb_2}$. The finite-time safe backward image of a set consists of all initial states from which trajectories, under a backup policy, remain in $\SSS_\rms$ for a finite horizon and reach the target set within that horizon. $\phi_{u_{\rmb_1}}$ and $\phi_{u_{\rmb_2}}$ demonstrate two of these trajectories corresponding to $u_{\rmb_1}$ and $u_{\rmb_2}$, respectively. The neural backup policy $u_\theta$ serves as an additional backup policy (i.e., $u_{\rmb_3} \equiv u_\theta$), trained using the RLBUS algorithm to expand the finite-time safe backward image of a subset of $\SSS_{\rmb_1} \cup \SSS_{\rmb_2}$. The set $S_{*_3}$ represents the finite-time safe backward image of a subset of this subset, while $\phi_{u_\theta}$ denotes sample trajecotries generated under $u_\theta$. The RLBUS algorithm ensures that $u_\theta$ is trained without any safety violations during training. It initiates exploration from the existing forward invariant subset of the safe set $\SSS_{*1} \cup \SSS_{*2}$ to achieve an expanded forward invariant set $\SSS_{*1} \cup \SSS_{*2} \cup \SSS_{*3}$.}
\label{fig:sets}
\end{figure}

One approach to ensuring safety with actuator constraints is the Backup Control Barrier Function (BCBF), which involves predicting system trajectories over a finite horizon to obtain a control-forward invariant subset of the safe set. For instance, \citep{gurriet2020scalable} determines this subset by constructing a CBF from finite-horizon predictions under a single backup control. Extending this approach, \citep{rabiee2023softmin,rabiee2023soft} introduced the soft-maximum/soft-minimum BCBF method, which leverages multiple backup controllers to further expand the forward-invariant subset. However, the resulting set remains conservative due to the design of the backup controllers

This paper introduces the Reinforcement Learning Backup Shield (RLBUS), a framework that guarantees safe exploration in RL by integrating BCBFs. RLBUS begins with a conservative forward-invariant set, constructed from multiple backup sets, which ensures safety under input constraints. It then employs model-free RL as an additional backup policy that train iteratively. This optimization enlarges the forward-invariant set, allowing RLBUS to explore a broader state space and enhance performance, all while guaranteeing zero safety violations throughout training. The effectiveness of RLBUS is demonstrated on an inverted pendulum example, where the expanded invariant set enables safe exploration across a broader state space, achieving improved performance without compromising safety.
% \footnote{Due to space limitations, the supplementary materials can be found at the end of the following version: \\ \href{https://arxiv.org/pdf/2312.07828}{https://arxiv.org/pdf/2312.07828}}. 
\section{Notation}

Let $\eta:[0,\infty) \times \BBR^n \to \BBR$ be continuously differentiable. 
Then, the partial Lie derivative of $\eta$ with respect to $x$ along the vector fields of $\psi:\mathbb{R}^n \to \mathbb{R}^{n \times \ell}$ is define as $
L_\psi \eta(t,x) \triangleq \frac{\partial \eta(t,x)}{\partial x} \psi(x)$.
Let $\mbox{int }\SA$, $\mbox{bd }\SA$ denote the interior and boundary of the set $\SA$. 
% Let $\SC^1$ denote the set of continuously differentiable functions. 
For a positive integer $n$, let $\BBI[n]$ denote $\{1, 2, \ldots, n\}$, and $\BBW[n]$ denote $\{0, 1, \ldots, n\}$.
% TODO: yechize behtar ja \chi bezarim
Let  $\epsilon \in \BBR \setminus \{0\}$, and let $\chi:\BBR^n\to \BBR$ be a continuously differentiable function. Define the set $\SB \triangleq \{ x \in \mathbb{R}^n : \chi(x) \geq 0 \}$. The $\epsilon$-superlevel set of the set $\SB$ is denoted by $\SB^\epsilon$ and is defined as $\SB^\epsilon \triangleq \{ x \in \mathbb{R}^n : \chi(x) \geq \epsilon \}$.

Let $\rho > 0$, and consider $\mbox{softmin}_\rho \colon \mathbb{R}^N \to \mathbb{R}$ and $\mbox{softmax}_\rho \colon \mathbb{R}^N \to \mathbb{R}$, which are defined by  $\mbox{softmin}_\rho (z_1,\cdots,z_N) \triangleq -\frac{1}{\rho}\log\sum_{i=1}^N e^{-\rho z_i}$ and $\mbox{softmax}_\rho (z_1,\cdots,z_N) \triangleq \frac{1}{\rho}\log\sum_{i=1}^N e^{\rho z_i} - \frac{\log N}{\rho}$, respectively. 
% The next result relates soft minimum and soft maximum to the minimum and maximum.
% \begin{aproposition} \label{prop:softmin_bound}
% \rm{
% Let $z_1,\ldots, z_N \in \BBR$. 
% Then,
% \begin{align*}
%  \min \, \{ z_1,\ldots, z_N \} - \frac{\log N }{\rho} 
%  \leq \mbox{softmin}_\rho(z_1,\ldots,z_N) \leq \min \, \{z_1,\ldots, z_N\},
% \end{align*}
% and
% \begin{align*} 
%  \max\,\{z_1,\ldots, z_N\}  - \frac{\log N}{\rho}
%  \leq \mbox{softmax}_\rho(z_1,\ldots,z_N)
%  \leq \max \, \{z_1,\ldots, z_N\}.
% \end{align*}
% }
% \end{aproposition}

\section{Problem Formulation}
Consider 
\begin{equation}\label{eq:dynamics}
\dot x(t) =\tilde f_u(x(t)) \triangleq f(x(t))+g(x(t)) u(x(t)),
\end{equation}
where $x(t) \in \BBR^{n}$ is the state, $x(0) = x_0 \in \BBR^n$ is the initial condition, $f: \BBR^n \to \BBR^n$ and $g: \BBR^n \to \BBR^{n \times m}$ are continuously differentiable on $\BBR^n$, $u:\BBR^n \to \SU \subseteq \BBR^m$ is the control, which is locally Lipschitz continuous on $\BBR^n$ and $\SU$ is defined as a compact and convex set of admissible controls, and $\tilde{f}_u: \BBR^n \to \BBR^n$ denotes the closed-loop dynamic under the control law $u$. We assume that the closed-loop dynamics $\tilde{f}_u$ is forward complete, meaning that the solution $x$ to~\eqref{eq:dynamics} exists for all $t \geq 0$.

Next, let $\phi_{u}: \BBR^n \times [0, \infty) \to \BBR^n$ defined by
\begin{equation}\label{eq:phi_def}    
    \phi_{u}(x, \tau) \triangleq x +\int_{0}^{\tau} \tilde f_{u}(\phi_u(x,\sigma)) \, \rmd \sigma,
\end{equation}
which is the \textit{flow map} of \eqref{eq:dynamics} under the control law $u$ with initial condition $x$ over a time period $\tau \in [0, \infty)$. The following definition is needed.

\begin{adefinition}\citep{rabiee2024closed}[Definition 1]\label{def:fwd_inv}\rm{
A set $\SD\subset \BBR^n$ is \textit{control forward invariant} with respect to \eqref{eq:dynamics}, if there exists a locally Lipschitz $\hat u:\SD \to \SU$ such that for all $x_0 \in \SD$ and all $t \geq 0$, $\phi_{u}(x_0,t)\in\SD$ with $u\equiv \hat u$}.
\end{adefinition}

Next, consider a continuously differentiable function $h_\rms \colon \BBR^n \to \BBR$, and define the safe set
\begin{equation}\label{eq:S_s}
\SSS_\rms \triangleq \{x \in \BBR^n \colon h_\rms(x) \geq 0 \},
\end{equation}
which comprises states that satisfy safety constraints. While $\SSS_\rms$ characterizes safety requirements, it is not necessarily control forward invariant. To ensure safety, it is necessary to identify a subset of $\SSS_\rms$ that is control forward invariant, and to remain in this set. However, if this identified subset is too small, it may adversely affect the performance of the task at hand. The performance is typically characterized by how closely an implemented control matches a desired performance policy. Let $u_\rmd: \BBR^n \to \SU$ be a performance policy designed to achieve control objectives without considering safety constraints. 
The key challenge is to identify a large enough control forward invariant subset of $\SSS_\rms$ that enables tracking $u_\rmd$ while guaranteeing safety.

To address this challenge, first, we assume the existence of multiple, relatively small, control forward invariant subsets of $\SSS_\rms$.
Formally, let $\ell$ be a positive integer and for each $j \in \BBI[\ell]$, let $h_{\rmb_j} \colon \BBR^n \to \BBR$ be continuously differentiable and define the $j$-th \textit{backup set}
\begin{equation}\label{eq:S_b_j}
\SSS_{\rmb_j} \triangleq \{x \in \BBR^n \colon h_{\rmb_j}(x) \geq 0 \},    
\end{equation}
and we assume $\SSS_{\rmb_j} \subseteq \SSS_\rms$. Next, for each $j \in \BBI [\ell]$, let $u_{\rmb_j}: \BBR^n \to \mathcal{U}$ be the $j$-th backup policy and make the following assumption.

\begin{aassumption}\label{assum:ub} \rm{
    For all $j \in \BBI[\ell]$, $x_0 \in \SSS_{\rmb_j}$ and $t \geq 0$, $\phi_{u_{\rmb_j}}(x_0,t) \in \SSS_{\rmb_j}$.
    }
\end{aassumption}

\Cref{assum:ub} implies that each backup policy $u_{\rmb_j}$ renders its corresponding backup set $\SSS_{\rmb_j}$ control forward invariant. The backup sets $\SSS_{\rmb_j}$ are small subsets of $\SSS_\rms$ that provide infinite-horizon safety guarantees and can typically be constructed using Lyapunov stability theory as a region of attraction around the equilibrium points inside $\SSS_\rms$.

To identify a larger control forward invariant subset of the safe set, a finite-time reach-avoid set problem is formulated. Let $T >0$ be a time horizon and
$\mathcal{U}^{[0,T]}$ denotes the set of measurable control functions mapping $[0,T]$ to the control space $\SU$ and consider the value function $V \colon \BBR^n \to \BBR$ defined by
\begin{equation}\label{eq:value function}
V(x) \triangleq \max_{\hat{u}(\cdot) \in \mathcal{U}^{[0,T]}} H(x,\hat{u}(\cdot)),
\end{equation}
where
\begin{equation}\label{eq:H_def}
    H(x, \hat u(\cdot)) \triangleq \min \left\{ \min_{\tau \in [0, T]} h_\rms(\phi_{\hat u(\cdot)}(x,\tau)), \max_{j \in \BBI[\ell]}h_{\rmb_j}(\phi_{\hat u(\cdot)}(x,T)) \right\},
\end{equation}
and define 
\begin{equation}\label{eq:R_def}
    \SR \triangleq \{ x \in \BBR^n \colon V(x) \geq 0 \}, 
\end{equation}
as the \textit{finite-time safe backward image} of $\bar \SSS_\rmb\triangleq \cup_{i \in \BBI[\ell]} \SSS_{\rmb_i}$. 
The set $\SR$ comprises all states $x \in \SSS_\rms$ for which a trajectory exists that (i) remains within the safe set $\SSS_\rms$ throughout $[0, T]$, and (ii) reaches at least one backup set $\SSS_{\rmb_j}$ at time $T$, and based on these properties, the set $\SR$ is control forward invariant.
This formulation represents a finite-time reach-avoid problem, with $\bar \SSS_\rmb$ as the reachable set and $\BBR^n \setminus \SSS_\rms$ as the avoid set. 
The objective is to identify the control forward invariant set $\SR\subseteq \SSS_\rms$. The size of this control forward invariant subset of $\SSS_\rms$ typically grows with the time horizon $T$, potentially providing sufficient space for tracking $u_\rmd$ while maintaining safety guarantees.

% \section{RLBUS: Reinforcement Learning Backup Control Barrier Function Shield}\label{sec:multibackup}

% \begin{aremark}
%     The backup control barrier function (BCBF) is a special case of the function $H(x; u(\cdot))$ where $\ell = 1$ and $u(\cdot)\equiv u_{\rmb}$. For the states inside $\SSS_{\rmb}$, using the backup control $u_{\rmb}$ ensures that the backup set $\SSS_{\rmb}$ remains control forward invariant. However, for states in $\SSS_\rms \setminus \SSS_{\rmb}$, using the backup control $u_{\rmb}$ may not be the optimal policy to reach the backup set and can result in a more conservative control forward invariant set. 
% \end{aremark}

\section{Proposed Method} \label{sec:method}

Traditional Hamilton-Jacobi reachability analysis transforms the optimization problem~\eqref{eq:value function} into a partial differential equation (PDE) using the Hamilton-Jacobi-Bellman principle, yielding a Hamilton-Jacobi-Isaacs equation with value function as a viscosity solution \citep{fisac2015reach}. However, this approach suffers from the curse of dimensionality, as solving the PDE on a discretized state-space grid becomes computationally intractable with increasing state-space dimension.
Moreover, given a specific task, obtaining the value function for all states is often unnecessary and computationally prohibitive.

To address these limitations, we propose to employ a learning-based method for approximating the value function. Consider a neural policy $u_\theta: \BBR^n \to \SU$ parameterized by $\theta \in \Theta \subseteq \BBR^d$, where $d$ represents the number of parameters. The objective is to train the neural policy $u_\theta$ to approximate the solution to \eqref{eq:value function}, while ensuring that safety is maintained throughout all training and execution episodes.

For $K$ training iterations, let
\begin{equation}\label{eq:value function neural}
V^{(k)}(x) \triangleq H(x, u_{\theta^{(k)}}), \quad \SR^{(k)} \triangleq \{ x \in \BBR^n \colon V^{(k)}(x) \geq 0 \},
\end{equation}
where $k \in \BBI[K]$ represents the iteration number and $\theta^{(k)}$ denotes the parameters at iteration $k$. While $V^{(k)}$ only approximates the value function $V$, the set $\SR^{(k)}$ maintains control forward invariance. This is because $\SR^{(k)}$ comprises all states where the flow of~\eqref{eq:dynamics} under $u_\theta^{(k)}$ remains in $\SSS_\rms$ for $t\in[0, T]$ and reaches $\bar \SSS_\rmb$. Moreover, as $V^{(k)}$ better approximates $V$, the set $\SR^{(k)}$ converges to $\SR$.

Unlike HJ reachability analysis which requires sweeping the entire state space, our proposed learning-based approach collects data guided by the performance policy $u_\rmd$. However, since $u_\rmd$ does not account for safety, we employ a minimum-intervention safety filtering approach: the system follows $u_\rmd$ unless safety is compromised, in which case it switches to the closest safe control to $u_\rmd$. To this end, in \Cref{sec:safe neural backup}, we propose a safety filtering approach based on backup control barrier functions~\citep{gurriet2020scalable, rabiee2023soft} with the neural policy serving as a backup policy. Notably, while the neural policy may perform poorly at the start of training, our framework guarantees that this does not compromise safety. Using this safety filter, safety is maintained during training, and as the neural policy improves, the size of the safely explorable region grows.
The training algorithm for the neural policy is detailed in \Cref{sec:RLBUS}.

\subsection{Safe Neural Backup Policy}\label{sec:safe neural backup}

This subsection develops a safety-guaranteed framework for neural backup policies. We first introduce a neural policy architecture that ensures safety even during early training stages when the neural policy may be unreliable. Next, we adopt the approach from \citep{rabiee2023soft,rabiee2023softmin} to synthesize a barrier function using the neural backup policy and the designed backup policies $\{u_{\rmb_1}, \ldots, u_{\rmb_\ell}\}$. Our proposed method improves the computational efficiency of \citep{rabiee2023soft}, and by leveraging the neural backup policy achieves a significantly larger forward invariant set compared to \citep{rabiee2023soft}.

Let $\nu > 0$ and consider a continuously differentiable function $\xi:\BBR \to [0,1]$ such that for all $x \in (-\infty,-\nu]$, $\xi(x) =0$; for all $x \in [0,\infty)$, $\xi(x)=1$; and $\xi$ is strictly increasing on $x \in [-\nu,0]$. Next, define $J(x) \triangleq \{j \in \BBI[\ell]: h_{\rmb_j}(x) \ge -\nu \}$, 
and make the following assumption.
% $\SSS_{\rmb_j}^{-\upsilon}  \triangleq \{ x \in \BBR^n : h_{\rmb_j}(x) \ge -a\}$
\begin{aassumption}\label{assum:j_singleton}
    \rm{
    $J(x)$ is singleton.
    }
\end{aassumption}

\Cref{assum:j_singleton} implies that for all $j_1, j_2 \in \mathbb{I}[\ell]$ with $j_1 \neq j_2$, $\SSS_{\rmb_{j_1}}^{-\nu} \cap \SSS_{\rmb_{j_2}}^{-\nu} = \emptyset$, ensuring that the $(-\nu)$-superlevel sets of the backup sets are disjoint.

Next, let $\pi_\theta \colon \BBR^n \to \SU$ be a multi-layer perceptron (MLP) parameterized by $\theta \in \Theta$ with continuously differentiable activation functions (e.g., ELU, SiLU). Then, we consider the following structure for the neural policy $u_\theta$.

\begin{equation}\label{eq:pi_shield}
    u_\theta(x) = \begin{cases}
            u_{\rmb_{J(x)}}(x),& \mbox{if } x \in \SSS_{\rmb_{J(x)}}, \\ 
            \xi(x)u_{\rmb_{_{J(x)}}}(x) + [1-\xi(x)] \pi_\theta(x), & \mbox{if } x \in \SSS_{\rmb_{J(x)}}^{-\nu} \setminus \SSS_{\rmb_{J(x)}},\\
            \pi_\theta(x),& \mbox{else}.
    \end{cases}
\end{equation}

The neural policy $u_\theta$ uses the MLP network for the states in $\SSS_\rms \setminus \SSS_{\rmb_{J(x)}}^{-\nu}$, and smoothly transitions to the backup control $u_{\rmb_{J(x)}}$ through the homotopy function $\xi$ as trajectories approach $\SSS_{\rmb_{J(x)}}$. 
Note that \Cref{assum:j_singleton} implies $u_\theta$ is well defined since $J(x)$ is unique.
The following result is immediately followed by the definition of $u_\theta$. The proof of the next result is in Section~\ref{sec:proofs}.
% \footnote{\href{https://arxiv.org/pdf/2312.07828}{https://arxiv.org/pdf/2312.07828}}.

\begin{aproposition}\label{prop:S_b_fwdinv}
\rm
Consider \eqref{eq:dynamics} where Assumptions~\ref{assum:ub} and~\ref{assum:j_singleton} are satisfied.
Let $x_0 \in \bar \SSS_\rmb$, then, for all $t\ge 0$, $\phi_u(x, t)\in \bar \SSS_\rmb$, with $u \equiv u_\theta$.
\end{aproposition}

Let $u_{\rmb_{\ell+1}} \equiv u_\theta$ be the $(\ell+1)$-th backup control. Let $\rho_1 >0$ and consider $h_{\rmb_{\ell+1}}: \BBR^n \to \BBR$ defined by
\begin{equation}\label{eq:backup_set_rl}
h_{\rmb_{\ell+1}}(x) \triangleq \mbox{softmax}_{\rho_1}(h_{\rmb_1}(x), h_{\rmb_2}(x), \ldots,h_{\rmb_\ell}(x)), 
\end{equation}
and define $\SSS_{\rmb_{\ell+1}} \triangleq \{x \in \BBR^n \colon h_{\rmb_{\ell+1}}(x) \geq 0 \}$ as the $(\ell+1)$-th backup set. The following result is the direct consequence of~\citep[Prop.~1]{rabiee2023soft}.
\begin{aproposition}\label{prop:bar_S_b}
$\SSS_{\rmb_{\ell+1}} \subseteq \bar \SSS_\rmb$.
\end{aproposition}
\Cref{prop:bar_S_b} implies that $\SSS_{\rmb_{\ell+1}}$ inner approximates $\bar \SSS_\rmb$, where $\rho_1$ determines the approximation accuracy.
The next result is the direct consequence of consequence of \Cref{prop:S_b_fwdinv} and \Cref{prop:bar_S_b}.
\begin{aproposition}\label{prop:S_b_new_fwdinv} \rm
Consider~\eqref{eq:dynamics} where Assumptions~\ref{assum:ub} and~\ref{assum:j_singleton} are satisfied. Let $x_0 \in \SSS_{\rmb_{\ell+1}}$, then, for all $t\ge 0$, $\phi_u(x, t)\in \bar \SSS_\rmb$, with $u \equiv u_\theta$.
\end{aproposition}

\Cref{prop:S_b_new_fwdinv} states that any trajectory initialized in $\SSS_{\rmb_{\ell+1}}$ under the neural policy $u_\theta$ remains in $\bar \SSS_\rmb$ for all time.

Next, for all $j \in \mathbb{I}[\ell+1]$, let $\tilde f_{u_{\rmb_j}} \colon \mathbb{R}^n \to \mathbb{R}^n$ be defined by \eqref{eq:dynamics} where $u$ is replaced by $u_{\rmb_j}$, and let $\phi_{u_{\rmb_j}} \colon \mathbb{R}^n \times [0, \infty) \to \mathbb{R}^n$ be defined by \eqref{eq:phi_def} where $\phi_u$ and $\tilde f_u$ are replaced by $\phi_{u_{\rmb_j}}$ and $\tilde f_{u_{\rmb_j}}$. 
In the following, we adopt the soft-maximum/soft-minimum barrier function method from~\citep{rabiee2023soft}, modifying the safe control formulation in~\eqref{eq:def_ua} and \eqref{eq:u_softmax_softmin} to achieve a simpler control structure.

Consider $h_{*j}, h_* \colon \BBR^n \to \BBR$ and $\SSS_*$ defined by 
\vspace{-0.5em}
\begin{gather}
h_{*j} \triangleq \min \, \left \{h_{{\rmb}_j}(\phi_{u_{\rmb_j}}(x,T)), \min_{\tau \in [0,T]} h_\rms(\phi_{u_{\rmb_j}}(x,\tau))\right \}, \quad \SSS_{*j} \triangleq \{ x \in \BBR^n \colon h_{*j}(x) \ge 0 \}\label{eq:h_min_cont_j_def}\\
h_*(x) \triangleq \max_{j\in\BBI[\ell+1]} h_{*j},\quad \SSS_* \triangleq \{ x \in \BBR^n \colon h_*(x) \ge 0 \} \label{eq:h_*_def},
\end{gather}
and it follows from~\citep[Prop.~8]{rabiee2023soft} that $\SSS_*$ is control forward invariant. $\SSS_*$ comprises all states for which there exists an index $\jmath\in \BBI[\ell+1]$ such that the trajectory under backup policy $u_{\rmb_\jmath}$ remains in $\SSS_\rms$ throughout $[0,T]$ and reaches $\SSS_{\rmb_\jmath}$ at time $T$. However, neither $h_{*j}$ nor $h$ can serve directly as barrier functions: $h_{*j}$ contains an embedded optimization problem and both functions lack differentiability. Instead, following \citep{rabiee2023soft}, we use a continuous approximation that evaluates $h_\rms$ along the backup policy trajectories at sampled time instances. Let $N$ be a positive integer, and define $T_\rms \triangleq T/N$. Let $\rho_2,\rho_3 > 0$ and for $j\in\BBI[\ell+1]$ consider $h_j, h:\BBR^n \to \BBR$ defined by
\begin{gather}
    h_j(x) \triangleq \mbox{softmin}_{\rho_2} ( h_\rms(\phi_{u_{\rmb_j}}(x,0)), h_\rms(\phi_{u_{\rmb_j}}(x,T_\rms)), \ldots, h_\rms(\phi_{u_{\rmb_j}}(x,NT_\rms)), h_\rmb(\phi_{u_{\rmb_j}}(x,NT_\rms))),\label{eq:sofmin_h_j_def} \\
    h(x) \triangleq \mbox{softmax}_{\rho_3} (h_1(x), \ldots, h_{\ell+1}(x)),\label{eq:h_softmaxmin_def}
 \end{gather}
and define
\begin{align}
\SSS_j &\triangleq \{x \in \BBR^n \colon h_j(x) \ge 0 \},\label{eq:def_S_j}\\
\SSS &\triangleq \{ x \in \BBR^n \colon h(x) \ge 0 \}\label{eq:def_S}.
\end{align}

Define $l_\phi \triangleq \max_{j \in \BBI[\ell+1]}\sup_{x\in \BBR^n} \| \tilde f_{u_{\rmb_j}}(x) \|_2$, and let $l_\rms$ denotes the Lipschitz constant of $h_\rms$ with respect to the two norm. Next, define $\epsilon_\rms \triangleq \frac{1}{2}T_\rms l_\phi l_\rms$. The follwoing results are needed. The first result is similar to~\citep[Prop. 10]{rabiee2023soft} and shows that a level set of $h$ is contained in $\SSS_*$. The second result relates $\SSS_{\ell+1}$ to $\SR$, and is directly followed from~\eqref{eq:H_def} and~\eqref{eq:h_softmaxmin_def}. The proof of the next results are in Section~\ref{sec:proofs}.

\begin{aproposition}\label{prop:S_eps}
    $\SSS^{\epsilon_\rms}\subseteq \SSS_*$.
\end{aproposition}

\begin{aproposition}\label{prop:R}
$\SSS_{\ell+1}^{\epsilon_\rms}\subseteq \SR$.
\end{aproposition}

Let $\alpha >0$, $\epsilon\in [0, \max_{x\in\SSS} h(x))$, and consider $\beta\colon \BBR^n \to \BBR$ defined by
\begin{equation}
\label{eq:feas_check}
\beta(x) \triangleq  L_f h(x) + \alpha (h(x)- \epsilon) + \max_{\hat u \in \SU} 
 L_g h(x)  \hat u,
\end{equation}
where for all $x \in \BBR^n$, $\beta(x)$ exists since $\SU$ is not empty.
Let $\kappa_h, \kappa_\beta > 0$ and consider $\gamma \colon \BBR^n \to \BBR$ defined by $\gamma(x) \triangleq \min \left\{\frac{h(x) - \epsilon}{\kappa_h}, \frac{\beta(x)}{\kappa_\beta} \right\}$, and define $\Gamma \triangleq \{ x \in\BBR^n \colon \gamma(x) \ge 0 \}$. For all $x \in \Gamma$, define
\vspace{-0.5em}
\begin{equation}
u_*(x) \triangleq \underset{\hat u \in \SU}{\mbox{argmin}}  \, 
\|\hat u - u_\rmd(x)\|_2^2 
\quad \text{subject to} \quad L_f h(x) + L_g h(x) \hat u + \alpha (h(x) - \epsilon) \ge 0. \label{eq:qp_softmin}
\end{equation} 
It follows from \citep[Prop.~6]{rabiee2023soft} that for all $x \in \Gamma$, the quadratic program \eqref{eq:qp_softmin} is feasible.
Next for all $x \in \BBR^n$, define $I(x) \triangleq \{j\in\BBI[\ell+1]: h_j(x) \ge \epsilon\}$, and $\bar \SSS \triangleq \cup_{j \in \BBI[\ell+1]}\SSS_j^\epsilon$. Then, for all $x \in \mbox{int } \bar \SSS$, define the augmented backup control
\begin{equation}\label{eq:def_ua}
u_\rma(x) \triangleq \dfrac{\sum_{j \in I(x)} [ h_j (x) - \epsilon] u_{\rmb_j}(x)}{\sum_{j \in I(x)} [h_j(x) - \epsilon]},
\end{equation}
which is a weighted sum of the backup controls for which $h_j(x) > \epsilon$. 
Note that for all $x \in \mbox{int }\bar \SSS$, $I(x)$ is not empty and thus, $u_\rma(x)$ is well defined.

Consider a continuous function $\sigma:\BBR \to [0,1]$ such that for all $a \in (-\infty,0]$, $\sigma(a) =0$; for all $a \in [1,\infty)$, $\sigma(a) =1$; and $\sigma$ is strictly increasing on $a \in [0,1]$. 

Finally, consider the control 
\begin{equation}\label{eq:u_softmax_softmin}
u(x) = \begin{cases} 
[1-\sigma(\gamma(x))] u_\rma(x) + \sigma(\gamma(x)) u_*(x),& \mbox{if } x \in \mbox{int } \Gamma, \\ 
%
% u_\rma(x), & \mbox{if } x \in \mbox{int } \bar \SSS_*^\epsilon \backslash \Gamma,\\
%
u_{\rmb_q}(x),& \mbox{else},
\end{cases}
\end{equation}
where $q:[0,\infty) \to \BBI[\ell+1]$ satisfies 
\begin{equation}\label{eq:def_q}
\begin{cases} 
\dot q = 0, & \mbox{if } x \not \in \mbox{bd } \Gamma,\\
q^+ \in I(x), & \mbox{if } x \in \mbox{bd } \Gamma,
\end{cases}
\end{equation}
where $q(0) \in \mathbb{I}[\ell+1]$ and $q^+$ denotes the value of $q$ after the instantaneous change. It follows from \eqref{eq:def_q} that $q$ remains constant when $x \not \in \bar \SSS$, and consequently, the same backup control $u_{\rmb_q}$ is used in \eqref{eq:u_softmax_softmin} until state $x$ reaches $\mbox{bd } \bar \SSS$. Thus, backup control switching occurs only on $\mbox{bd } \bar \SSS$. The controls~\eqref{eq:def_ua} and \eqref{eq:u_softmax_softmin} are modified versions of those in~\citep{rabiee2023soft}, offering a simpler formulation. While this modification leads to earlier switching to backup policies, the effect becomes negligible for larger values of $\rho_2$ and $\rho_3$, which is typical in practice.

The following theorem which is similar to~\citep[Theorem 2]{rabiee2023soft} establishes that the control law $u$ defined by~\eqref{eq:sofmin_h_j_def}--\eqref{eq:def_q} is continuous and admissible, and renders $\SSS_*$ forward invariant. The proof of the theorem is necessary due to the changes in~\eqref{eq:sofmin_h_j_def}--\eqref{eq:def_q}. However, the proof follows the same logic as that of~\citep[Theorem 2]{rabiee2023soft} and is provided in Section~\ref{sec:proofs}.

\begin{atheorem}\label{thm:softmax_softmin}
\rm 
Consider \eqref{eq:dynamics}, where Assumptions~\ref{assum:ub} and~\ref{assum:j_singleton} are satisfied and $u$ is given by~\eqref{eq:sofmin_h_j_def}--\eqref{eq:def_q}.
Then, the following conditions hold:
\begin{enumerate}[label=\textit{(\alph*)}]
\item 
$u$ is continuous on $\BBR^n \setminus \mbox{bd } \Gamma$.
\vspace{-0.5em}
\label{thm:softmax_u_continuity}
\item 
For all $x \in \BBR^n$, $u(x) \in \SU$. 
\vspace{-0.5em}
\label{thm:softmax_u_cond}
\item 
\label{thm:softmax_forward_inv}
Let $\epsilon \ge \epsilon_\rms$, $x_0 \in \SSS_*$, and $q(0) \in \{ j \colon h_{*_j}(x_0) \ge 0 \}$.
Then, for all $t \ge 0$, $x(t) \in \SSS_* \subseteq \SSS_\rms$.
\end{enumerate}
\end{atheorem}

To summarize, we provided a structure for the neural policy to be used as a backup policy by~\eqref{eq:pi_shield} and introduced the backup set corresponding to this backup policy in~\eqref{eq:backup_set_rl}. We also demonstrated how this learning-based backup policy can be used alongside other predesigned backup policies. Furthermore, we showed that even if the neural backup policy performs arbitrarily poorly at the beginning of training, the safety guarantees established by Theorem~\ref{thm:softmax_softmin} remain valid.

The control~\eqref{eq:u_softmax_softmin} relies on computing the Lie derivatives $L_fh(x)$ and $L_gh(x)$, which are utilized in~\eqref{eq:feas_check} and~\eqref{eq:qp_softmin}. In \citep{rabiee2023soft,gurriet2020scalable}, these Lie derivatives are computed through sensitivity analysis, where a sensitivity matrix $Q \in \mathbb{R}^{n \times n}$ captures the evolution of state trajectory variations with respect to initial conditions. This approach requires solving an augmented system of $n + n^2$ ordinary differential equations (ODEs). While this method ensures exact computation of the Lie derivatives, it introduces significant computational overhead due to the quadratic growth in the number of equations with respect to the state dimension.

To overcome this computational burden, this paper instead leverages the adjoint sensitivity method introduced by \cite{chen2018neural}. The adjoint method introduces an adjoint state $a(t) \in \mathbb{R}^n$ that evolves backward according to $da/dt = -a(t)^\top \partial \tilde f_{u_{\rmb_j}}/\partial x$. This approach offers significant computational advantages by reducing the dimensionality of the augmented system from $n + n^2$ to $2n$. Furthermore, unlike the forward sensitivity approach which requires storing the entire state trajectory, the adjoint method can reconstruct necessary forward states during the backward pass, achieving $O(1)$ memory complexity with respect to the integration time steps. This enables efficient computation of the Lie derivatives for real-time control applications.

\cref{alg:backupshield} outlines the procedure for implementing the control defined in~\eqref{eq:sofmin_h_j_def}--\eqref{eq:def_q}.

% man vpspace furu kardam 2 kaht munde... ino age mituni bekah yekam. manzuram one yekam kholasash kon
% \begin{aremark}\rm
%     The desired control $u_d$ in the quadratic program~\eqref{eq:qp_softmin} can be replaced by a neural desired policy $u_{d_{\tilde\theta}}:\mathbb{R}^n \to \BBR^m$ parameterized by $\tilde\theta \in \Theta$. This creates a dual-network framework where $u_{d_{\tilde\theta}}$ and the neural backup policy $u_\theta$ are trained concurrently while maintaining safety guarantees. As $u_{d_{\tilde\theta}}$ guides state trajectories toward high-performance regions, it concentrates the training of $u_\theta$ in performance-critical areas, establishing a performance-oriented framework with guaranteed safety.
% \end{aremark}

\begin{aremark}\rm
The desired control $u_d$ in the quadratic program~\eqref{eq:qp_softmin} can be replaced by a neural desired policy $u_{d_{\tilde\theta}}:\mathbb{R}^n \to \BBR^m$ parameterized by $\tilde\theta \in \Theta$, which is trained to optimize performance objectives. This substitution introduces a dual-network framework where the neural desired policy $u_{d_{\tilde\theta}}$ and the neural backup policy $u_\theta$ are trained concurrently while ensuring safety guarantees throughout the learning process. Specifically, $u_{d_{\tilde\theta}}$ steers the state trajectories toward high-performance regions, inherently focusing the training data for $ u_\theta $ on these critical regions. This targeted data collection enables $u_\theta$ to refine backup policies in areas of frequent operation, thereby expanding the forward-invariant set in these regions. This cooperative learning between the policies establishes a performance-oriented framework where safety is guaranteed throughout the training.
\end{aremark}

\subsection{RLBUS: Reinforcement Learning Backup Control Barrier Function Shield}\label{sec:RLBUS}

This section presents a method to train the neural policy $u_\theta$ using the reinforcement learning (RL) framework.
We consider the standard RL setup with a finite-horizon deterministic Markov decision process (MDP) defined by tuple $(\mathcal{\SX}, \mathcal{\SU}, \gamma, r, \mu_0, H, f, g)$, where $\SX\in \BBR^n$ is the state space, $\SU$ is the action space, $r: \SX \times \SU \rightarrow \BBR$ is the reward function, $\gamma\in(0,1)$ is the discount factor, $\mu_0$ is the distribution of the initial state $x_0$, $H>0$ is the horizon, and $f, g$ as in~\eqref{eq:dynamics} represent the deterministic dynamics model. Let $\delta t >0$, and note that the state transitions for the MDP are obtained by zero-order-holding \eqref{eq:dynamics} on the control $u$ for $\delta t$. 
We note that this approximation does not affect safety guarantees. Provided $\delta t$ is sufficiently small, and under some Lipschitz-continuity assumptions solving the constrained optimization in \eqref{eq:qp_softmin} at the required frequency ensures safety between triggering times, as highlighted in \citep[Theorem 3]{ames2016control}. Specifically, with appropriate choices of $\delta t$ and $\epsilon$ in~\eqref{eq:qp_softmin}, safety constraints are satisfied for all times.

Let $\mu^{u_\theta}$ denote the distribution of states and actions induced by policy $u_\theta$. Define the expected discounted return $\eta:\SU\to\BBR$ as
\begin{equation}
    \eta(u_\theta) = \BBE_{\mu^{u_\theta}} \left[\sum_{i=1}^{H} \gamma^{i-1} r\left(x(t), u_\theta(x(t))\right)\right].
\end{equation}

The policy $u_\theta$ can be trained to maximize $\eta(u_\theta)$ using standard RL methods. In what follows, we provide details on the reward structure and data collection process.
For notational simplicity, let $x_{j,i}$ represent the $i$-th sampled point along the trajectory under the backup policy $u_{\mathrm{b}_j}$, starting from state $x$, i.e., $x_{j,i}=\phi_{u_{\rmb_j}}(x,iT_\rms)$. During data collection at state $x$, for each $j\in \BBI[\ell+1]$:
\vspace{-0.5em}
\begin{enumerate}[label=(\textit{\arabic*})]
    \item Generate the trajectory under backup policy $u_{\rmb_j}$ to obtain $\left\{x_{j,i}\right\}_{i=0}^N$
    \vspace{-0.5em}

    \item Compute $h_j(x)$ from~\eqref{eq:h_softmaxmin_def}.
    \vspace{-0.5em}

    \item Assign the reward
    \begin{equation}\label{eq:rl_backup_reward}
        r_j(x_{j,i}, u_{\rmb_j}(x_{j,i})) \triangleq h_j(x), \quad \text{for all $i\in \BBI[N]$}.
    \end{equation}
    The reward function~\eqref{eq:rl_backup_reward} essentially assigns the same reward to all the state-action pairs along the trajectory under backup policy $u_{\rmb_j}$.
    \vspace{-0.5em}

    \item Add the set of tuples $\left\{(x_{j,i}, u_{\rmb_j}(x_{j,i}), x_{j,i+1}, r_j)\right\}_{i=0}^N$ to the replay buffer.
\end{enumerate}

Note that for $j = \ell+1$, corresponding to the neural backup policy $u_\theta$, the reward function matches the expression in \eqref{eq:H_def}, with the exception that the maximum function is replaced by the soft maximum, and $h_\rms$ is evaluated at sampled times. This makes the reward essentially equal to $h_{\ell+1}(x)$. This formulation directly inspired the reward function in~\eqref{eq:rl_backup_reward}. Although the trajectories under the neural backup policy $u_\theta$ alone could suffice for training the backup policy, we utilized the trajectories generated under all backup policies during safety filtering to augment the training dataset without incurring additional computation. Specifically, the reward for these trajectories is given by the value of $h_j$ for $j \in \mathbb{I}[\ell]$. By leveraging this reward, the policy is encouraged to maximize $h_j$ for the neural policy.
\cref{alg:rlbus} outlines the RLBUS algorithm.
\begin{figure}[t!]
    \centering
    \subfigure[Episode 0]{%
        \includegraphics[width=0.40\textwidth,clip=true,trim=0.1in 0.0in 0.5in 0.2in]{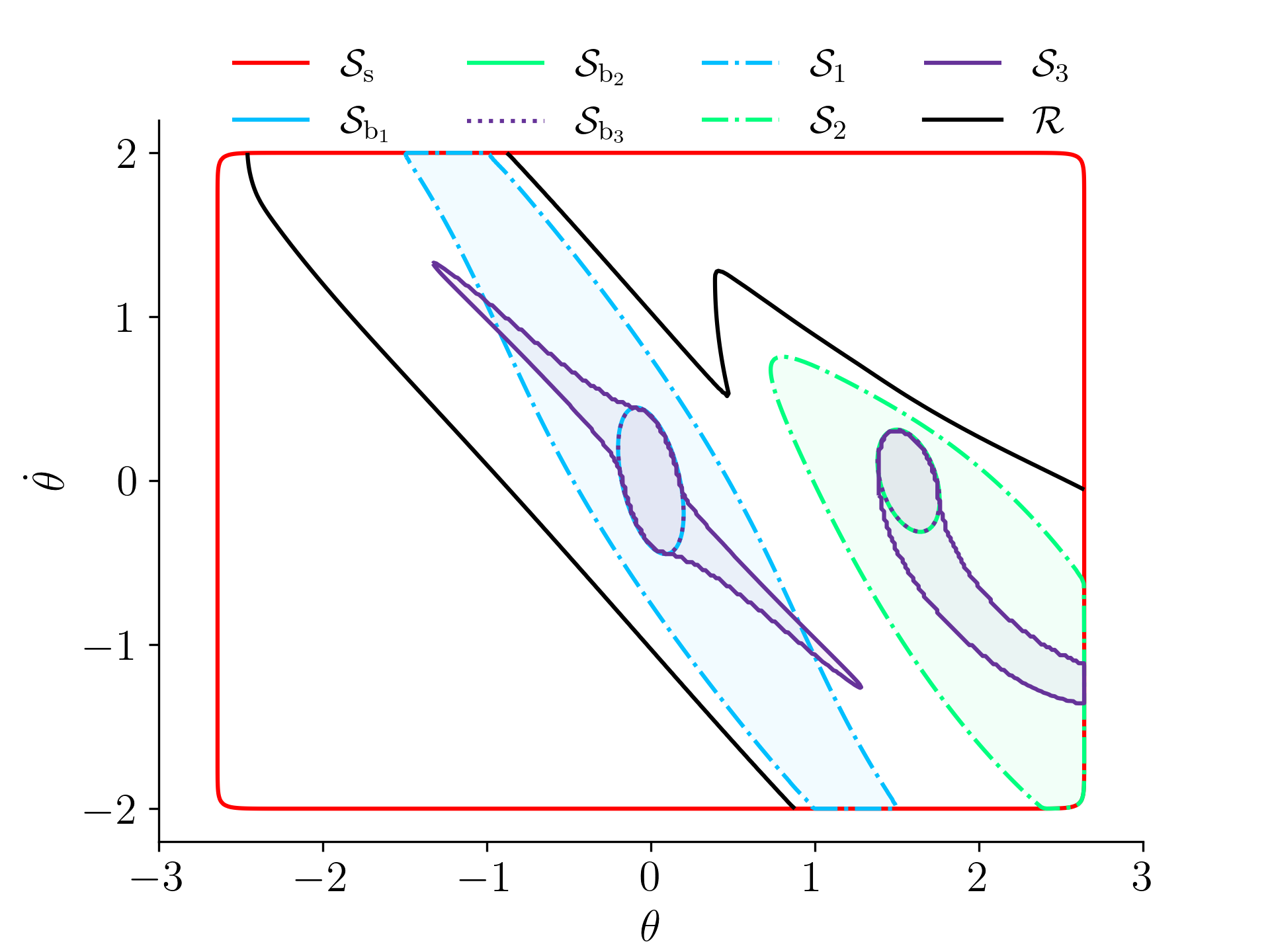}%
        \label{fig:set_exp_a}%
    }
    % \subfigure[Episode 5]{%
    %     \includegraphics[width=0.32\textwidth,clip=true,trim=0.1in 0.0in 0.5in 0.2in]{h_countor_2}%
    %     \label{fig:set_exp_mid}%
    % }
    \subfigure[Episode 100]{%
        \includegraphics[width=0.40\textwidth,clip=true,trim=0.1in 0.0in 0.5in 0.2in]{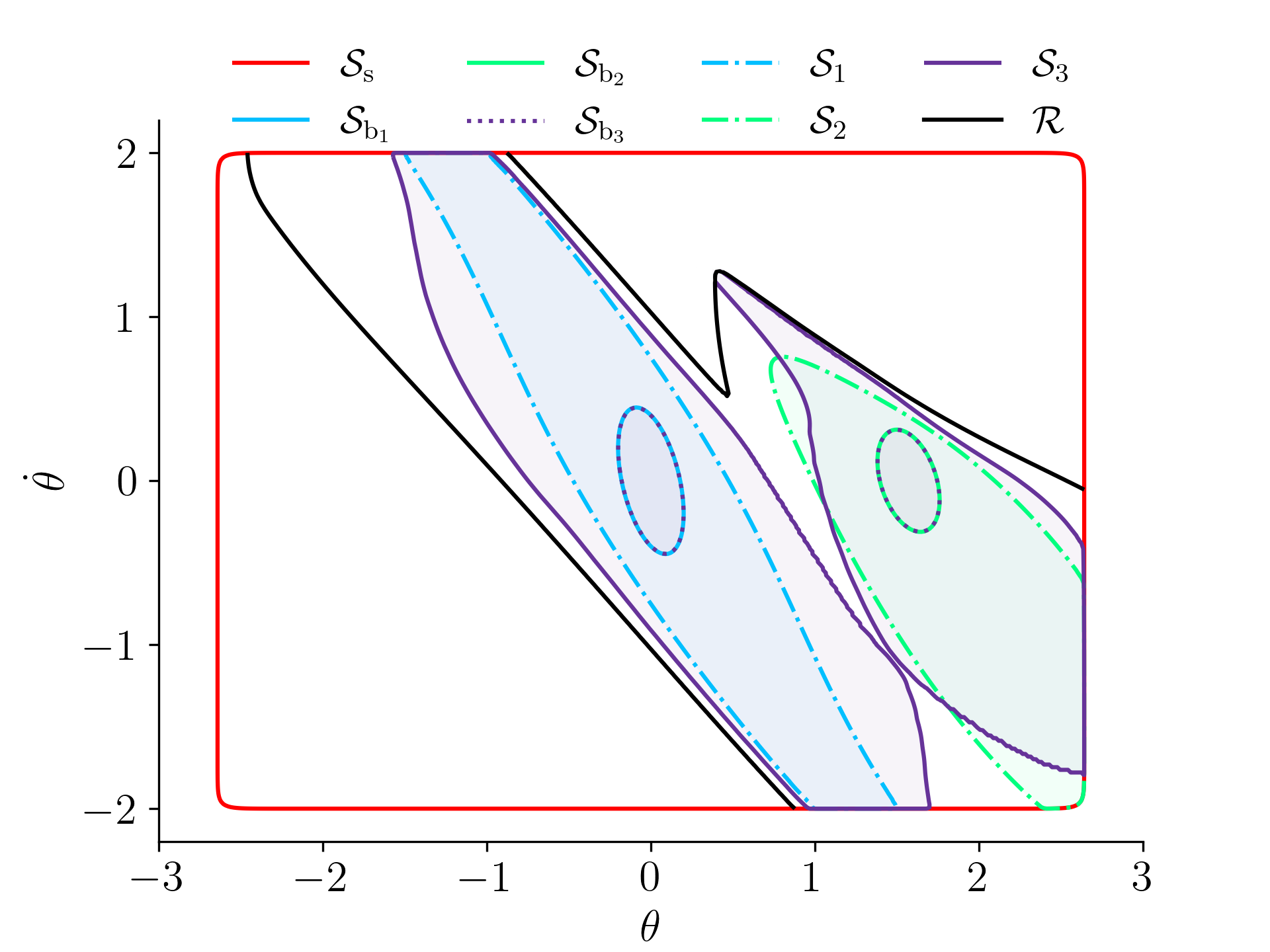}%
        \label{fig:set_exp_b}%
    }
    \caption{Illustration of $\SSS_\rms$, $\SSS_{\rmb_1}$, $\SSS_{\rmb_2}$, $\SSS_{\rmb_3}$, $\SSS_{1}$, $\SSS_{2}$, $\SSS_{2}$, and $\SR$.}
    \label{fig:set_exp}
\end{figure}

\section{Numerical Results}

\begin{figure}[t!]
    \centering
    \subfigure[Safety violation]{\includegraphics[width=0.45\textwidth,clip=true,trim= 0.1in 0.0in 0.0in 0in]{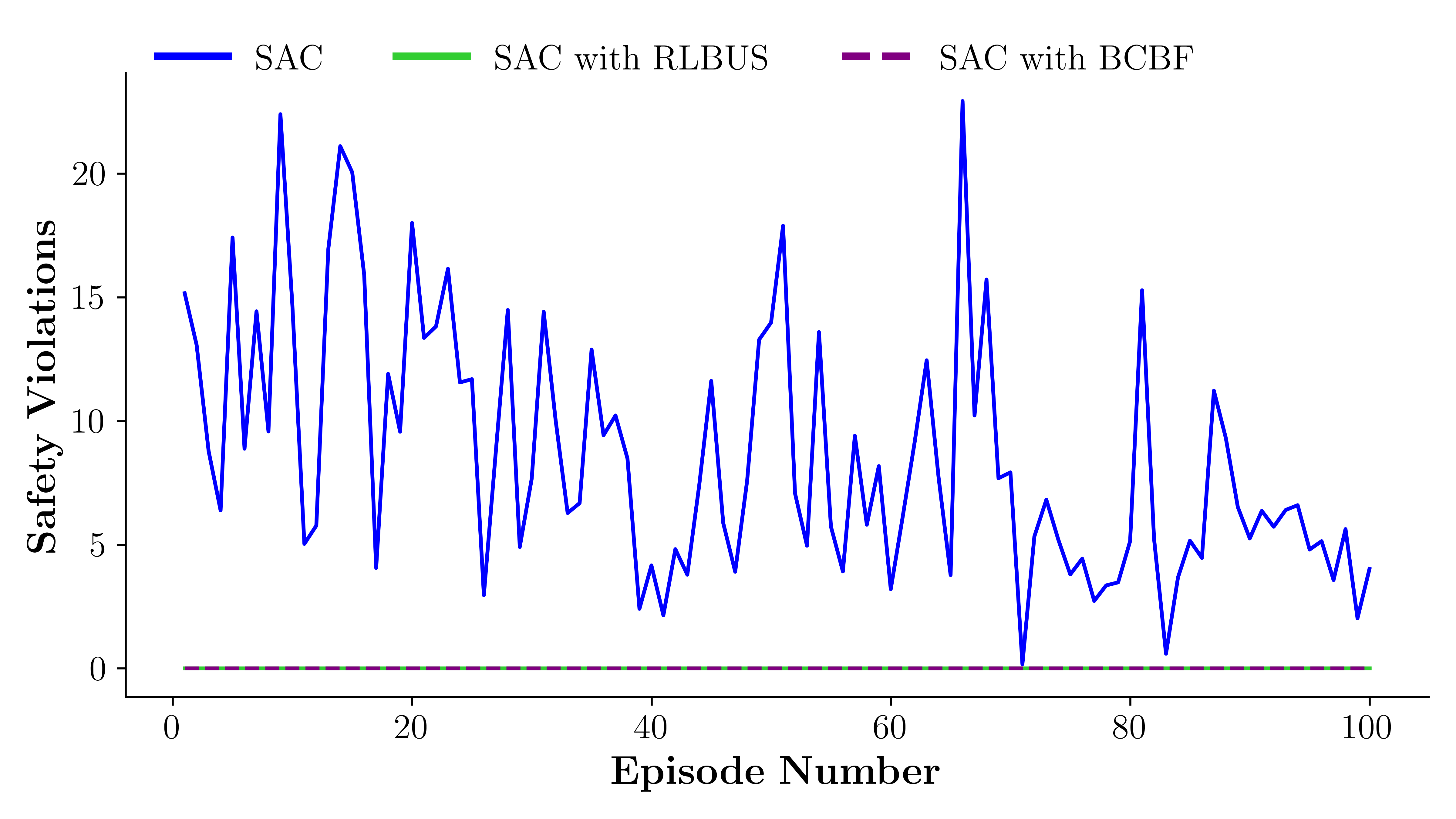} \label{fig:safety_violation}} 
    \subfigure[Sampling Return]{\includegraphics[width=0.45\textwidth,clip=true,trim= 0.1in 0.0in 0.0in 0.0in]{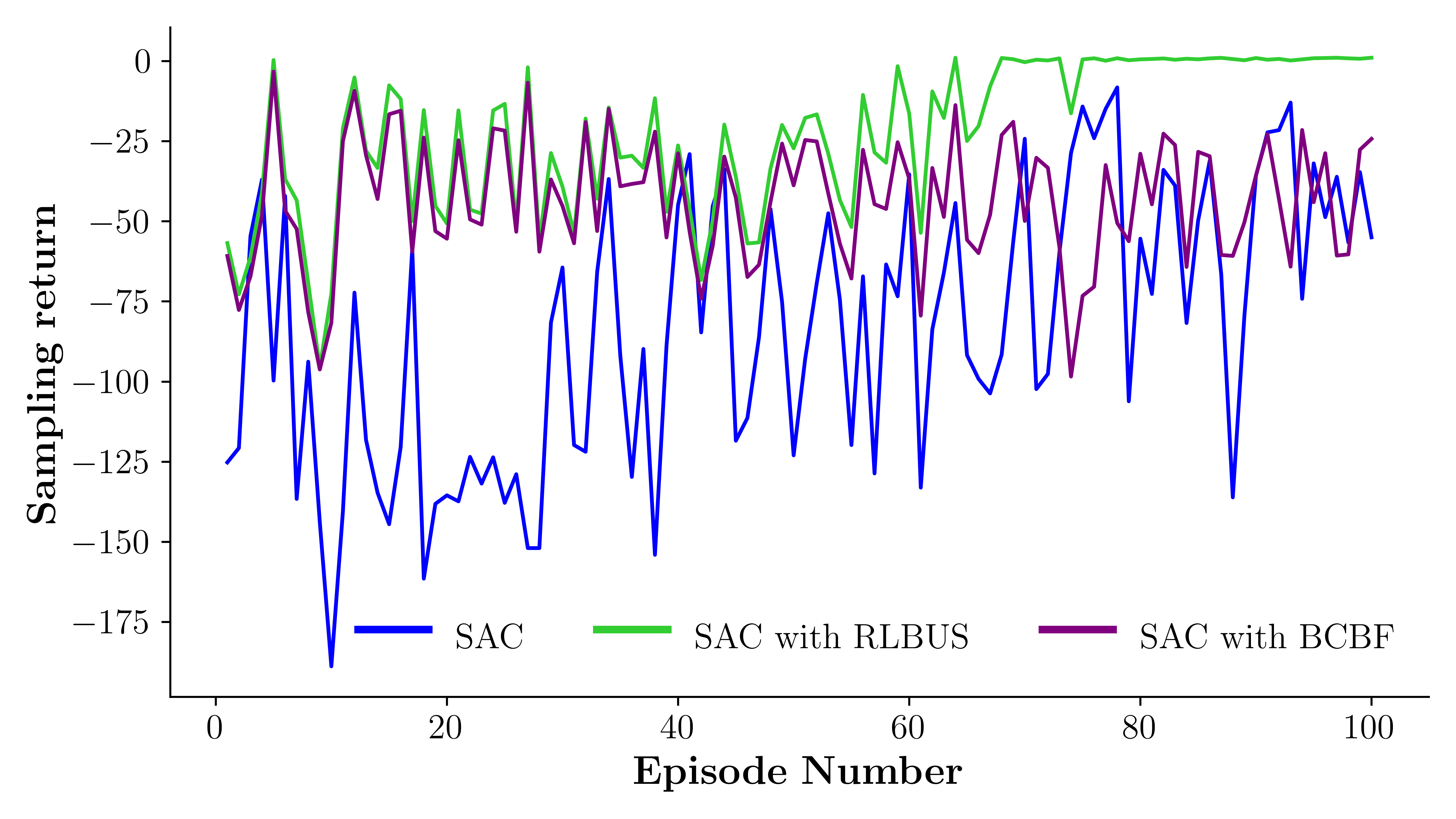} \label{fig:sac_reward}}     
    \caption{Safety violations and sampling returns of the performance agent across three scenarios: \textit{(i)} SAC, \textit{(ii)} SAC with BCBF, and \textit{(iii)} SAC with RLBUS.}
    \label{fig:ret_vio}
\end{figure}

% Pendulum Dynamics
% Pendulum desired control
Consider the inverted pendulum modeled by \eqref{eq:dynamics}, where 
\begin{equation*}
    f(x) = \begin{bmatrix}
    \dot \varphi \\
    \sin \varphi
    \end{bmatrix}, 
    \qquad
    g(x) = \begin{bmatrix}
    0 \\
    1
    \end{bmatrix}, 
    \qquad 
    x = \begin{bmatrix}
    \varphi \\
    \dot \varphi
    \end{bmatrix}, 
\end{equation*}
and $\varphi$ is the angle from the inverted equilibrium.
Let $\bar u = 1.5$ and $\SU = \{ u \in \BBR \colon u \in [-\bar u, \bar u] \}$.
The safe set $\SSS_\rms$ is given by \eqref{eq:S_s}, where $h_\rms(x) = 1 - \left \| \matls \frac{1}{\pi-0.5} & 0 \\ 0 & 0.5 \matrs x \right \|_{100}.$

We consider two backup sets. Specifically, for $j \in \BBI[2]$, The backup safe set $\SSS_{\rmb_j}$ is given by \eqref{eq:S_b_j}, where $h_{\rmb_j}(x) = 0.02 - (x - x_{\rmb_j})^\rmT Pj (x - x_{\rmb_j})$, $P_1 = \matls 0.625 &  0.125 \\ 0.125 & 0.125 \matrs$, and $P2 = \matls 0.650 &  0.150 \\ 0.150 & 0.240 \matrs$. Next, consider $u_{\rmb_j}(x) = \bar u \tanh \frac{1}{\bar u} K(x-x_{\rmb_j})$, where $x_{\rmb_1}\triangleq[\, 0\quad 0 \, ]^\rmT$, $x_{\rmb_2}\triangleq[\, {\pi}/{2}\quad 0 \, ]^\rmT$. Lyapunov's direct method can be used to confirm that Assumption \ref{assum:ub} is satisfied.
%Horizon yadam rafte. \rho inam adadesho nadadam
The neural backup policy $u_\theta$ is trained using the approach described in \Cref{sec:method}, aiming to maximize the reward~\eqref{eq:rl_backup_reward} to expand the identified control forward invariant subset of the safe set. 

\Cref{fig:set_exp} visualizes the sets $\SSS_\rms$, $\SSS_{\rmb_1}$, $\SSS_{\rmb_2}$, $\SSS_{\rmb_3}$, $\SSS_{1}$, $\SSS_{2}$, $\SSS_{3}$, and $\SR$ at two different episodes. The set $\SR$ is derived by solving \eqref{eq:value function} using the Hamilton-Jacobi-Bellman PDE, representing the largest finite-time safe backward image of $\bar \SSS_\rmb$. \Cref{fig:set_exp_a} shows the initial state of the set $\SSS_3$ at the start of the first episode, while \Cref{fig:set_exp_b} illustrates the expansion of $\SSS_3$ after 100 episodes of training the neural backup policy $u_\theta$.

To evaluate the performance and safety violations of the proposed framework, we consider training a neural desired policy $u_{d_{\tilde\theta}}:\mathbb{R}^2 \to \SU$ parameterized by $\tilde\theta \in \Theta$, trained as a RL agent to maximize the reward function $r_{\rmp}:\mathbb{R}^2 \times \SU \to \mathbb{R}$ defined by $r_{\rmp}(\varphi, \dot{\varphi}, u) \triangleq -(\varphi - 0.8)^2 - 0.1\dot{\varphi}^2 - 0.001u^2$.

Both the neural policy $u_\theta$ and the neural desired policy $u_{d_{\tilde\theta}}$ are trained using the soft actor-critic method~\citep{haarnoja2018soft}.
The reward $r_\rmp (\varphi, \dot \varphi, u)$ is maximized at $u = 0$ and $x_{\rm{opt}} = [0.8 \quad 0]^T$. However, this state lies outside of $\SSS_1 \cup \SSS_2$, and achieving optimal performance requires expanding the forward invariant set. We evaluate the framework under three different scenarios: \textit{(i)} SAC: training the neural desired policy without any safety constraints, \textit{(ii)} SAC with BCBF: training the neural desired policy with a BCBF safety constraint using two backup policies $u_{\rmb_1}$ and $u_{\rmb_2}$, and \textit{(iii)} SAC with RLBUS: concurrently training the neural desired policy and a neural backup policy within the proposed framework with backup policies $u_{\rmb_1}$, $u_{\rmb_2}$, and $u_\theta$.

\Cref{fig:ret_vio} compares safety violations and sampling returns across the three scenarios. \Cref{fig:safety_violation} shows that while scenario \textit{(i)} exhibits multiple safety violations, scenarios \textit{(ii)} and \textit{(iii)} maintain safety throughout the training process. Additionally, \Cref{fig:sac_reward} demonstrates that scenarios \textit{(iii)} achieve higher sampling returns compared to scenario \textit{(ii)}. Initially, the $x_{\rm{opt}}$ lies outside $\bigcup_{j \in \mathbb{I}[3]} \mathcal{S}_j $, making it infeasible to reach. However, \Cref{fig:set_exp} illustrates that as the forward invariant set expands, the optimal performance state becomes contained within $\mathcal{S}_3 $. This expansion allows scenario \textit{(iii)} to achieve higher sampling returns than scenario \textit{(ii)}. Together, \Cref{fig:set_exp} and \Cref{fig:ret_vio} demonstrate that the RLBUS expands the forward invariant set towards performance-optimal regions while maintaining safety throughout the training process.

\bibliography{Backup-RL.bib}

\newpage
\section{Supplementary Materials}\label{sec:supp_mat}
\subsection{Algorithms}

\begin{algorithm2e}[ht]
\small
\DontPrintSemicolon

% \LinesNumbered
\caption{BackupShield \citep{rabiee2023soft}}\label{alg:backupshield}
\SetKwData{Nhorizon}{$N$}
\SetKwData{FeasScale}{$\kappa_\beta$}
\SetKwData{hScale}{$\kappa_h$}
\SetKwData{Sigmafunc}{$\sigma$}
\SetKwComment{Comment}{$\triangleright$\ }{}
\Comment{\textcolor{blue}{Global:$h_\rms, h_{\rmb_j}, u_{\rmb_j}, u_\rmd, N, T_\rms, \sigma, \rho_1, \rho_2, \rho_3, \kappa_\beta, \kappa_h, \alpha, \epsilon, \pi_\theta$}}
\KwIn{$x$: state}
\KwOut{$u$, $\{h_j\}_{j=0}^{\ell + 1}$, $\Big\{\{\phi_{u_{\rmb_j}}(x,iT_\rms)\}_{i=0}^N\Big\}_{j=0}^{\ell +1}$} 
\SetKwFunction{BCBF}{BackupShield}
\SetKwProg{Fn}{Function}{:}{}
\Fn{\BCBF{$x$}}{
    % \Comment{\ProgSty{\textcolor{blue}{Add Neural Policy to backup policies}}}

    % $u_{\rmb_{\ell+1}} \gets$ \eqref{eq:pi_shield}\;
    % \Comment{\ProgSty{\textcolor{blue}{ADD SOME COMMENTS TO DIFFERENT STEPS OF THIS ALGORITHM}}}
    
    \For{$j=1,\ldots, \ell +1$}{
        Solve numerically to obtain $\{\phi_{u_{\rmb_j}}(x,iT_\rms)\}_{i=0}^N$\; 
        $h_j \gets$ \eqref{eq:sofmin_h_j_def}\;
    }
    $h \gets$ \eqref{eq:h_softmaxmin_def}\;
    $\beta \gets$\eqref{eq:feas_check}\;
    $\gamma \gets \min \{\frac{h - \epsilon}{\kappa_h}, \frac{\beta}{\kappa_\beta}\}$\; 
    % \Comment{\ProgSty{Compute importance sampling weights}}
    \eIf{$\gamma \le 0$}{
        $u \gets u_{\rmb_q}(x)$ where $q$ satisfies~\eqref{eq:def_q}
        }
    {
    Compute $L_fh(x)$ and $L_gh(x)$\;
    $u_\rma \gets$ \eqref{eq:def_ua}\;
    $u_* \gets$ solution to quadratic program~\eqref{eq:qp_softmin}\;
    $u \gets [1-\sigma(\gamma)] u_\rma + \sigma(\gamma) u_*$\;}
\Return{$u$, $\{h_j\}_{j=0}^{\ell + 1}$, $\Big\{\{\phi_{u_{\rmb_j}}(x,iT_\rms)\}_{i=0}^N\Big\}_{j=0}^{\ell +1}$}
}
\textbf{End Function}
\end{algorithm2e}

\newpage

\begin{algorithm2e}[ht]\label{alg:RLBUS}
\small
\DontPrintSemicolon
% \LinesNumbered
\caption{RLBUS: Reinforcement Learning Backup Shield}\label{alg:rlbus}
\SetKwComment{Comment}{$\triangleright$\ }{}
\Comment{\textcolor{blue}{Global:$\theta, \mu_0, \SD$\textrm{ (replay buffer), }$\delta t$\textrm{ (simulation sampling time), }$K$\textrm{ (maximum episode steps), }$T_\rmf$\textrm{ (training frequency)}}}
\SetKwFunction{BCBF}{BackupShield}
\SetKwProg{Fn}{Function}{:}{}
\For{$\text{episode}=1,2,\ldots$}{
\Comment{\ProgSty{\textcolor{blue}{Data collection}}}
$x \gets \mu_0.\FuncSty{\text sample()}$\;
    \For{$k=1,2\ldots, K$}{
        \Comment{\ProgSty{\textcolor{blue}{Obtain safe control and backup trajectories}}}
        $u, \{h_j\}_{j=0}^{\ell +1}, \Big\{\{\phi_{u_{\rmb_j}}(x,iT_\rms)\}_{i=0}^N\Big\}_{j=0}^{\ell +1} \gets \FuncSty{\text{BackupShield}}(x)$\;
        % \#\ProgSty{ Data Collection}\;
        \Comment{\ProgSty{\textcolor{blue}{Prepare training samples}}}
        \For{$j=1,\ldots,\ell +1$}{
            $r \gets h_j$\;
            \For{$i=1,\ldots,N$}{ 
            $\hat x \gets \phi_{u_{\rmb_j}}(x,iT_\rms)$\;
            $\hat u \gets u_{\rmb_j}(\hat x)$\; 
            \uIf{$i < N$}{
            $\hat x^\prime \gets \phi_{u_{\rmb_j}}(x, (i+1)T_\rms)$\;
            }
            \Else{
            \Comment{\ProgSty{\textcolor{blue}{Step forward from $\hat x$ for $T_\rms$ to obtain $\hat x^\prime$ at the $N$-th step}}}
            % \#\ProgSty{ step forward from $\hat x$ for $T_\rms$}\;
            $\hat x^\prime \gets \phi_{u_{\rmb_j}}(\hat x, T_\rms)$\;
            }
            \Comment{\ProgSty{\textcolor{blue}{Add data to buffer}}}
            $\SD$.\FuncSty{\text{append}(}$\hat x, \hat u, r, \hat x^\prime\FuncSty{)}$\;                        
        }        
        }
        \Comment{\ProgSty{\textcolor{blue}{Execute the safe control and step forward for one time step}}}
        $x \gets \phi_u(x, \delta t)$\;                     
}

        \Comment{\ProgSty{\textcolor{blue}{Training}}}
         \If{($episode \% T_\rmf)= 0$}
         {Update $\theta$ using samples from $\SD$}
}
\Return $\pi_{\theta}$
\end{algorithm2e}

\newpage
\subsection{Proofs of Propositions \ref{prop:S_b_fwdinv}, \ref{prop:S_eps}, \ref{prop:R} and \Cref{thm:softmax_softmin}}\label{sec:proofs}

\begin{aproof}[\noindent Proof of \Cref{prop:S_b_fwdinv}]
% Change j_* to J(x_0)
From \Cref{assum:j_singleton}, we know that for all $j_1, j_2 \in \mathbb{I}[\ell]$ with $j_1 \neq j_2$, $\SSS_{\rmb_{j_1}}^{-\nu} \cap \SSS_{\rmb_{j_2}}^{-\nu} = \emptyset$. Since $\SSS_{\rmb_{j_1}} \subset \SSS_{\rmb_{j_1}}^{-\nu}$ and $\SSS_{\rmb_{j_2}} \subset \SSS_{\rmb_{j_2}}^{-\nu}$ for all $j_1, j_2$, this implies that $\SSS_{\rmb_{j_1}} \cap \SSS_{\rmb_{j_2}} = \emptyset$. Given $x_0 \in \bar \SSS_\rmb = \bigcup_{j \in \BBI[\ell]} \SSS_{\rmb_j}$, define $j_* \triangleq \{j \in \mathbb{I}[\ell] : x_0 \in \SSS_{\rmb_j}\}$. From the disjointness of backup sets established above, $j_*$ must be singleton, i.e., $x_0$ belongs to exactly one backup set. 

Now, assume for contradiction that there exists $t_1 > 0$ such that $x(t_1) \notin \SSS_{\rmb_{j_*}}$ and for all $t < t_1$, $x(t) \in \SSS_{\rmb_{j_*}}$. Then, since for all $t < t_1$, $x(t) \in \SSS_{\rmb_{j_*}}$, from \eqref{eq:pi_shield} we have that $u_\theta(x(t)) = u_{\rmb_{j_*}}(x(t))$ for all $t < t_1$. Therefore, from \Cref{assum:ub}, we must have $x(t_1) \in \SSS_{\rmb_{j_*}}$, which contradicts our assumption. This proves that $x(t) \in \SSS_{\rmb_{j_*}} \subset \bar{\SSS}_\rmb$ for all $t \geq 0$.
\end{aproof}

% Let $x_1\in \bar \SSS_\rmb = \cup_{j \in \BBI[\ell]} \SSS_{\rmb_j}$.
% First, \Cref{assum:j_singleton} implies that for all $j_1, j_2 \in \mathbb{I}[\ell]$ with $j_1 \neq j_2$, $\SSS_{\rmb_{j_1}}^{-\nu} \cap \SSS_{\rmb_{j_2}}^{-\nu} = \emptyset$. This, together with the fact that  $\SSS_{\rmb_{j_1}}\subset\SSS_{\rmb_{j_1}}^{-\nu}$ and $\SSS_{\rmb_{j_2}}\subset\SSS_{\rmb_{j_2}}^{-\nu}$, implies that $\SSS_{\rmb_{j_1}}\cap \SSS_{\rmb_{j_2}} = \emptyset$.

% Thus, for all $x\in \bar \SSS_\rmb = \cup_{i \in \BBI[\ell]} \SSS_{\rmb_i}$, 

% Since $x_0 \in \bar \SSS_\rmb = \cup_{i \in \BBI[\ell]} \SSS_{\rmb_i}$
    
\vspace{1em}

\begin{aproof}[\noindent Proof of \Cref{prop:S_eps}]
Let $x_1 \in \SSS^{\epsilon_\rms}$ and let $j^* = \underset{j \in \BBI[\ell+1]}{\text{argmax}}, h_j(x_1)$. Let $\tau \in [0,T]$ and $i \in \BBI[N]$ be such that $|\tau - iT_\rms| < \frac{T_\rms}{2}$. Then, it follows from \citep[Lemma 1]{gurriet2020scalable} that
\begin{equation*}
    | h_\rms(\phi_{j_*}(x_1,\tau)) - h_\rms(\phi_{j_*}(x_1,i_*T_\rms)) | \, \le \epsilon_\rms,
\end{equation*}
which implies that
\begin{equation}\label{eq:h_s_tau_cond_j}
    h_\rms(\phi_{j_*}(x_1,\tau)) \ge h_\rms(\phi_{j_*}(x_1,i_*T_\rms)) -\epsilon_\rms.
\end{equation}
Next, \citep[Prop.~1]{rabiee2023soft} implies that $h_{j_*}(x_1) \geq h(x_1)$. Also, since $x_1 \in \SSS^{\epsilon_\rms}$, \eqref{eq:def_S} implies that $h(x_1) \geq \epsilon_\rms$. These together imply that $h_{j_*}(x_1) \geq \epsilon_\rms$. Thus,
\begin{equation}\label{eq:temp1}
  \min \left \{h_{\rmb_{j_*}}(\phi_{u_{\rmb_j}}(x_1,NT_\rms)), \min_{i \in \{0,1,\ldots,N\}} h_\rms(\phi_{u_{\rmb_{j_*}}}(x_1,iT_\rms)) \right \} \ge h_{j_*}(x_1),
\end{equation}
which implies that $h_\rms(\phi_{u_{\rmb_{j_*}}}(x_1,i_*T_\rms)) \ge \epsilon_\rms$.
Combining this with~\eqref{eq:h_s_tau_cond_j}, conclude that $h_\rms(\phi_{j_*}(x_1,\tau))\ge 0$, which implies
\begin{equation}\label{eq:temp3}
\min_{\tau \in [0,T]} h_\rms(\phi_{j_*}(x_1,\tau)) \ge 0.
\end{equation}

Next, since $h_{j_*}(x_1)\ge \epsilon_\rms$, from~\eqref{eq:temp1} we have
\begin{equation}\label{eq:temp2}
h_{\rmb_{j_*}}(\phi_{u_{\rmb_j}}(x_1,T)) = h_{\rmb_{j_*}}(\phi_{u_{\rmb_j}}(x_1,NT_\rms)) \ge \epsilon_\rms.
\end{equation}
Therefore, \eqref{eq:h_min_cont_j_def}, \eqref{eq:temp3}, and \eqref{eq:temp2} imply that
\begin{equation}\label{eq:temp5}
    h_{*{j_*}}(x_1) \ge 0.
\end{equation}
Finally, since $h_*(x_1) = \max_{j \in \BBI[\ell+1] } \, h_{*_j}(x_1) \ge h_{*_{j_*}}(x_1)$, combined with~\eqref{eq:temp5}, conclude that $h_*(x_1) \ge 0$, which implies $x_1 \in \SSS_*$. Thus, $ \SSS^{\epsilon_\rms} \subseteq \SSS_*$.
\end{aproof}
\vspace{1em}
\begin{aproof}[Proof of ~\Cref{prop:R}]
The following definitions are needed. Define
\begin{align*}
\hat H(x) \triangleq \min \bigg\{ &\min_{\tau \in [0, T]} h_\rms(\phi_{u_\theta}(x,\tau)),\\ 
&\mbox{softmax}_{\rho_1}\left(h_{\rmb_1}(\phi_{u_\theta}(x,T), h_{\rmb_2}(\phi_{u_\theta}(x,T)), \ldots, h_{\rmb_\ell}(\phi_{u_\theta}(x,T)\right) \bigg\},
\end{align*}
and
\begin{align*}
\hat H_\rms(x) \triangleq \min \bigg\{ &\min_{i \in \{0,1,\ldots, N\}} h_\rms(\phi_{u_\theta}(x,iT_\rms)), \\
&\mbox{softmax}_{\rho_1}(h_{\rmb_1}\left(\phi_{u_\theta}(x,T), h_{\rmb_2}(\phi_{u_\theta}(x,T)), \ldots, h_{\rmb_\ell}(\phi_{u_\theta}(x,T)\right) \bigg\}.
\end{align*}

Let $x_1 \in \BBR^n$. Then, from Proposition 1 in \citep{rabiee2023soft}, $\hat H(x_1) \le H(x_1, u_\theta(\cdot))$ and
$h_{\ell+1}(x_1) \le \hat H_{\rms}(x_1)$.
From Lemma 1 in \citep{gurriet2020scalable}, $\hat H_\rms(x_1) - \epsilon_\rms \le \hat H(x_1)$. Moreover, from~\eqref{eq:value function}, $H(x_1, u_\theta(\cdot)) \le V(x_1)$. Thus, 
\begin{equation}\label{eq:temp_prop5_1}
h_{\ell+1}(x_1) - \epsilon_\rms \le \hat H_\rms(x_1) - \epsilon_\rms \le \hat H(x_1)\le H(x_1, u_\theta(\cdot)) \le V(x_1).    
\end{equation}
Finally,~\eqref{eq:R_def}, \eqref{eq:def_S_j}, and \eqref{eq:temp_prop5_1} imply that $\SSS_{\ell+1}^{\epsilon_\rms} \subset \SR$.
\end{aproof}

\vspace{1em}
The following lemmas are needed for the proof of \Cref{thm:softmax_softmin}. Lemma~\ref{lem:u_a} is similar to Proposition 11 of \citep{rabiee2023soft}, and the proof follows a similar approach, with $\nu$, $\bar{h}_{*_j}$, and $\SSS_\epsilon$ replaced by $\ell+1$, $h_{*_j}$, and $\bar{\SSS}$, respectively. Lemma~\ref{lem:Gamma} is a consequence of Propositions 1 and 12 of~\citep{rabiee2023soft}.

\begin{lemma}\label{lem:u_a}\rm 
    $u_\rma$ is continuous on $ \mbox{int } \bar \SSS$.
\end{lemma}

\begin{lemma}\label{lem:Gamma}\rm
$\Gamma \subseteq \mbox{int }\SSS^\epsilon \subseteq \mbox{int } \bar \SSS$.
\end{lemma}
\begin{aproof}
First, from Proposition 1 of~\citep{rabiee2023soft} $\SSS^\epsilon \subseteq \bar \SSS$, which implies that $\mbox{int }\SSS^\epsilon \subseteq \mbox{int } \bar \SSS$. This together with the fact that from Proposition 12 of~\citep{rabiee2023soft}, $\Gamma \subseteq \mbox{int }\SSS^\epsilon$, imply that $\Gamma \subseteq \mbox{int } \SSS^\epsilon \subseteq \mbox{int } \bar \SSS$.
\end{aproof}

\vspace{1em}
\begin{aproof}[\noindent Proof of \Cref{thm:softmax_softmin}]

To prove~\ref{thm:softmax_u_continuity}, first the proof to Theorem 2 in~\citep{rabiee2023soft} imply that $u_*$ and $\gamma$ are continuous on $\Gamma$. Next, Lemma~\ref{lem:u_a} and Lemma~\ref{lem:Gamma} imply that $u_\rma$ is continuous on $\Gamma$.
Define $u_\rmm \triangleq [1-\sigma(\gamma(x))] u_\rma(x) + \sigma(\gamma(x)) u_*(x)$.
Since, in addition, $\sigma$ is continuous on $\BBR$, it follows that $u_\rmm$ continuous on $\mbox{int }\Gamma$.
Next, \eqref{eq:def_q} implies for $x \in \BBR^n \backslash \Gamma$, $q$ is constant.
Since, in addition, $u_{\rmb_j}$ is continuous on $\BBR^n$, it follows from \eqref{eq:u_softmax_softmin} that $u$ is continuous on $\BBR^n \backslash \Gamma$.
Thus, $u$ is continuous on $(\mbox{int } \Gamma) \cup ( \BBR^n \backslash \Gamma)$, which confirms~\ref{thm:softmax_u_continuity}.

To prove~\ref{thm:softmax_u_cond}, let $d\in \BBR^n$, and we consider 2 cases: $d\in \BBR^n \backslash \Gamma$, and $d \in \mbox{int } \Gamma$. First, let $d\in \BBR^n \backslash \Gamma$, and \eqref{eq:u_softmax_softmin} implies $u(d) = u_{\rmb_q}(d)\in\SU$.
Next, let $d \in \Gamma$.
Since for $j\in\{1,\ldots,\ell+1\}$, $u_{\rmb_j}(d) \in\SU$, it follows from~\eqref{eq:def_ua} that $u_\rma(d) \in \SU$. 
Since $u_\rma(d),u_*(d) \in \SU$, the same arguments in the proof to Theorem 1 of \citep{rabiee2023soft} with $u_\rmb$ replaced by $u_\rma$ imply that $u(d) \in \SU$, which confirms~\ref{thm:softmax_u_cond}.

To prove~\ref{thm:softmax_forward_inv},
since $\epsilon \ge \epsilon_\rms$, it follows from~\Cref{prop:S_eps} and~\Cref{lem:Gamma} that $\Gamma \subset \SSS^\epsilon \subseteq \SSS_*$. 
Define $\SSS_{*j}  \triangleq \{ x \in \BBR^n :h_{*j}(x) \ge 0 \}$, and it follows from Lemma 1 in~\citep{gurriet2020scalable} that $\SSS_j^\epsilon \subseteq \SSS_{*j}$.

Let $t_1 \ge 0$, and assume for contradiction that $x(t_1) \not \in \SSS_*$, which implies  $x(t_1) \not \in \Gamma$.
Next, we consider two cases: (i) there exists $t \in [0,t_1)$ such that $x(t) \in \Gamma$, and (ii) for all $t \in [0,t_1)$, $x(t) \not \in \Gamma$. 

First, consider case (i), and it follows that there exists $t_2 \in [0,t_1)$ such that $x(t_2) \in \mbox{bd } \Gamma$ and for all $\tau \in (t_1,t_2]$, $x(\tau) \not \in \Gamma$. 
Thus, \eqref{eq:def_q} and \eqref{eq:u_softmax_softmin} imply that there exists $\lambda \in I(x(t_2))$ such that for all $\tau \in [t_2,t_3]$, $q(\tau) = \lambda$ and  $u(x(\tau)) = u_{\rmb_\lambda}(x(\tau))$. Since, in addition, $x(t_2) \in \SSS_\lambda^{\epsilon} \subseteq \SSS_{*_\lambda}$, it follows from Proposition 8 of~\citep{rabiee2023soft} that $x(t_2) \in \SSS_*$, which is a contradiction. 

Next, consider case (ii), and \eqref{eq:u_softmax_softmin} and \eqref{eq:def_q} imply that for all $\tau \in [0,t_2]$, $q(\tau) = q(0)$ and $u(x(\tau)) = u_{\rmb_{q(0)}}(x(\tau))$. 
Since, in addition, $x_0 \in \SSS_{*_{q(0)}}$, Proposition 8 of~\citep{rabiee2023soft} implies $x(t_2) \in \SSS_*$, which is a contradiction. 

\end{aproof}

\end{document}